\documentclass[12pt,a4paper]{article}
\usepackage{color,graphicx}
\usepackage{eufrak}

\def\lvec#1{\setbox0=\hbox{$#1$}
    \setbox1=\hbox{$\scriptstyle\leftarrow$}
    #1\kern-\wd0\smash{
    \raise\ht0\hbox{$\raise1pt\hbox{$\scriptstyle\leftarrow$}$}}
    \kern-\wd1\kern\wd0}
\def\rvec#1{\setbox0=\hbox{$#1$}
    \setbox1=\hbox{$\scriptstyle\rightarrow$}
    #1\kern-\wd0\smash{
    \raise\ht0\hbox{$\raise1pt\hbox{$\scriptstyle\rightarrow$}$}}
    \kern-\wd1\kern\wd0}
\def\diracstar#1#2{
    \setbox0=\hbox{$\gamma$}\setbox1=\hbox{$\gamma_{#1}$}
    \gamma_{#1}\kern-\wd1\kern\wd0
    \smash{\raise4.5pt\hbox{$\scriptstyle#2$}}}
\newcommand{\beq}{\begin{equation}}
\newcommand{\eeq}{\end{equation}}
\newcommand{\beqn}{\begin{eqnarray}}
\newcommand{\eeqn}{\end{eqnarray}}
\newcommand{\nn}{\nonumber}

\begin{document}

\begin{titlepage}
\pagestyle{empty}
\date{}
\title{
Euclidean versus Minkowski short distance
\vspace*{3mm}}

\author{
G.C.\ Rossi$^{a)\,b)}$  \, M.\ Testa$^{c)}$
}
\maketitle
\begin{center}
  $^{a)}${\small Dipartimento di Fisica, Universit\`a di  Roma
  ``{\it Tor Vergata}'' \\ 
  INFN, Sezione di Roma 2}\\
  {\small Via della Ricerca Scientifica - 00133 Roma, Italy}\\
  \vspace{.2cm}
 $^{b)}${\small Centro Fermi - Museo Storico della Fisica e Centro Studi e Ricerche 
 ``E.\ Fermi''}\\
  {\small Piazza del Viminale 1 - 00184 Roma, Italy}\\
    \vspace{.2cm}
    $^{c)}${\small Dipartimento di Fisica, Universit\`a di  Roma
  ``{\it La Sapienza}'' \\ 
  INFN, Sezione di Roma  ``{\it La Sapienza}''}\\
    {\small Piazzale A.\ Moro 5 - 00187 Roma, Italy}\\
\end{center}

\abstract{In this note we reexamine the possibility of extracting parton distribution functions from lattice simulations. We discuss the case of quasi-parton distribution functions, the possibility of using the reduced Ioffe-time distributions and the more recent proposal of directly making reference to the computation of the current-current $T$-product. We show that in all cases the process of renormalization hindered by lattice momenta limitation represents an obstruction to a direct Euclidean calculation of the parton distribution function.}

\end{titlepage}
\newpage

\section{Introduction}
\label{sec:INTRO}

After the publication of the paper of ref.~\cite{Ji:2013dva} a substantial amount of work has been invested in the attempt of computing parton distribution functions (PDFs) from first principle lattice simulations~\footnote{In this short note we cannot give due credit to all the Authors working in this field for lack of space. For a useful list of references one can look at the recently published PDF-white-paper~\cite{Lin:2017snn}.}. 

The possibility of extracting PDFs from lattice QCD simulations of the (matched) hadronic matrix elements of a bilocal operator~\footnote{We wish to recall that the notion of bilocal in this context was first introduced in ref.~\cite{Brandt:1972nw}.} has been analyzed critically in ref.~\cite{Rossi:2017muf}, where it was observed that, despite the fact that such lattice-derived quantities can be made UV finite by an appropriate multiplicative renormalization, moments are plagued by UV power divergences, so that the resummed expression provided by lattice simulations does not yield the correct PDF, as the moments of the physical PDF are instead finite and experimentally measured quantities. 

In spite of these difficulties a large number of papers have appeared which addressed the (perturbative) calculation of matching and renormalization coefficients, report lattice data of matrix elements of bilocal operators and enlarge the set of lattice quantities that can be possibly used for the purpose of extracting PDFs.

In this note we start in sect.~\ref{sec:JI} by reexamining the theoretical foundation of the Ji proposal of ref.~\cite{Ji:2013dva} strengthening the argument given in~\cite{Rossi:2017muf}. In sect.~\ref{sec:RIT} we extend the discussion to the case in which the reduced Ioffe-time distributions are considered~\cite{Orginos:2017kos,Zhang:2018ggy}. In sect.~\ref{sec:QIUMA} we illustrate the problems associated with the proposal of extracting the PDF's from the lattice hadronic matrix element of the $T$-product of two currents advocated in~\cite{Ma:2017pxb}. We end in sect.~\ref{sec:CONCL} with a few remarks. 

\section{The Ji approach}
\label{sec:JI} 

We start the discussion by considering the unrealistic situation in which the theory is canonical. We then describe the modifications occurring when it is not, separately analyzing the Minkowski and Euclidean case. We conclude that the problem of getting UV finite moments represents an obstruction to a direct lattice calculation of PDF's starting from the hadronic matrix elements of the Ji bilocal operator.  

\subsection{Minkowski metrics}
\label{sec:MM}

With reference to the simplified situation of a scalar current $J(\xi) = \phi^2 (\xi)$, also discussed in~\cite{Rossi:2017muf}, one can write for the hadronic deep inelastic cross section in the parton approximation 
\begin{eqnarray}
\hspace{-.8cm}&&(2 \pi)^4 W (q^2, q\cdot P) = \int d^4 \xi e^{- i q \cdot \xi} \langle P| \phi (0) \phi(\xi)| P\rangle  \Delta (\xi) = \nonumber \\
 \hspace{-.8cm}&&=  \sum_n \int \frac {d {\bf k}} {2 |{\bf k}|}|\langle n|\phi(0)|P\rangle |^2 (2 \pi)^4 \delta^4 (P+q-p_n -k) \, , 
 \label{deepin}
\end{eqnarray}
where
\begin{eqnarray}
\Delta (\xi) \equiv \int \frac {d {\bf k}} {2 |{\bf k}|} e^{i k \cdot \xi} = \int d^4 k \, \delta (k^2) \theta (k^0)\, e^{ik\cdot \xi} 
\end{eqnarray}
and $k^\mu \equiv (|{\bf k}|, {\bf k})$ is the massless parton final momentum. The ket $|P\rangle$ denotes a covariantly normalized hadron state with four-momentum $P$.

Lorentz invariance implies
\begin{eqnarray}
\langle P|\phi (0) \phi(\xi)|P\rangle = {\cal M}(P \cdot \xi, \xi^2) \, . 
\label{lorbil}
\end{eqnarray}
In the canonical case ${\cal M}(P \cdot \xi, \xi^2)$ is a regular function that needs to be evaluated for $\xi^2 \approx 0$. We are interested in the computation of its Fourier Transform (FT)
\begin{eqnarray}
&& {\cal M}(P \cdot \xi, 0) = \int_{- \infty}^{+ \infty} d x f (x) \, e^{- i x P\cdot \xi} \label{bilocstru1} \\
&&f (x) = \frac {1}{2 \pi} \int_{- \infty}^{+ \infty} {\cal M}(P \cdot \xi, 0) \,e^{i x P\cdot \xi} d (P\cdot \xi) \, , \label{bilocstru}
\end{eqnarray} 
as $f(x)$ is related to $W(q^2, q\cdot P)$ by
\begin{eqnarray}
&&(2 \pi)^4 W (q^2, q\cdot P) = \int_{- \infty}^{+ \infty} d x f (x) \int d^4 \xi \, e^{- i (q + x P) \cdot \xi} \Delta (\xi) = \nonumber \\
&&= (2 \pi)^4 \int_{- \infty}^{+ \infty} d x f (x) \delta [(q + x P)^2] \theta [ (q + x P)^0] \, ,
\end{eqnarray}
finally leading in the canonical Bjorken limit to the relation   
\begin{eqnarray}
W (q^2, q\cdot P) \approx \frac {x f (x)}{-q^2} \, , \qquad x= \frac{-q^2}{2q\cdot P} \, .\label{structure}
\end{eqnarray}
This is the standard argument formally relating the structure function (i.e.\ the FT of the bilocal matrix element~(\ref{lorbil})) to the deep inelastic cross section. 

In eqs.~(\ref{bilocstru1}) and~(\ref{bilocstru}) the bilocal operator can be Taylor expanded around $\xi=0$, yielding 
\begin{eqnarray}
\hspace{-.8cm}&&\langle P| \phi (0) \phi(\xi) |P \rangle \!= \!\sum_{n=0}^{\infty} \frac {1} {n!} \langle P|\phi(0) \frac{\partial}{\partial \xi^{\mu_1}}\frac{\partial}{\partial \xi^{\mu_2}}\ldots \frac{\partial}{\partial \xi^{\mu_n}}\phi (\xi)\Big{|}_{\xi=0}|P\rangle \xi^{\mu_1} \xi^{\mu_2} \dots \xi^{\mu_n} \!\equiv \!\nonumber\\
\hspace{-.8cm}&&\equiv \sum_{n=0}^{\infty} \langle P| O_{\mu_1\mu_2 \dots \mu_n} (0) |P \rangle \xi^{\mu_1} \xi^{\mu_2} \dots \xi^{\mu_n} \, .
\end{eqnarray}
The matrix elements of $O_{\mu_1 \mu_2\dots \mu_n} (0)$ are of the form
\begin{eqnarray}
\langle P| O_{\mu_1 \mu_2\dots \mu_n} (0) |P \rangle = A_n P_{\mu_1}P_{\mu_2} \dots P_{\mu_n} + traces \, , \label{locopbil}
\end{eqnarray}
where $traces$ denote form factors containing some $g_{\mu_i \mu_j}$ tensor. For example, in the case of $O_{\mu_1 \mu_2} (0)$, we have
\begin{eqnarray}
\langle P| O_{\mu_1 \mu_2} (0) |P \rangle = A_2 P_{\mu_1} P_{\mu_2} + B_2 g_{\mu_1 \mu_2} \, .
\end{eqnarray}
The physical PDFs are related to the $A_n$ form factors (moments), while the $traces$ $B_n$ are spurious contributions which need to be subtracted out. In the Minkowski region this subtraction is automatically performed by taking $\xi^2 = 0$ (as in eq.~(\ref{bilocstru1})). In the Euclidean case the situation is more complicated.

\subsection{Euclidean metrics}
\label{sec:EM}

Eliminating trace terms is problematic in the case in which only Euclidean data are available, as it happens in lattice computations. In this case the only available information making direct contact with Minkowski physics are the matrix elements of the bilocal operator $\phi (0) \phi (\xi)$ at equal times ($\xi= (0,0,0,z)$). Eq.~(\ref{lorbil}) is still valid and in this kinematical situation reads 
\begin{eqnarray}
\langle P|\phi (0) \phi(z)|P\rangle = {\cal M}(-P_z \, z, -z^2) \, . \label{lorbile}
\end{eqnarray}
In Euclidean metrics, in order to eliminate the trace terms we can take advantage of the fact that the bilocal matrix element~(\ref{lorbile}) is a function of two independent variables, which may be chosen to be $\nu \equiv - P_z\,z$ (the so-called Ioffe time~\cite{IOFFE}) and $\beta \equiv - z^2$, so that one can recover the required structure function from the formula
\begin{eqnarray}
f (x) = \lim_{\beta \to0} \frac {1}{2 \pi} \int_{- \infty}^{+ \infty} {\cal M} (\nu, \beta)\, e^{i x \nu} d \nu \, .\label{tracecan}
\end{eqnarray} 
Eq.~(\ref{tracecan})~\footnote{At non-vanishing $\beta$ the FT in the r.h.s.\ of eq.~(\ref{tracecan}) defines what in the literature is called the {\it pseudo}-PDF~\cite{Radyushkin:2017cyf}.}  shows that in order to remove the trace terms in Euclidean region we must know $\langle P|\phi (0) \phi(z)|P\rangle$ for $P_z \to \infty$ as $z\to 0$, while keeping  $\nu=-P_z \, z$ fixed. In lattice simulations this requirement poses a serious problem as momenta are bounded from above by the inverse lattice spacing, $a^{-1}$, which in turn limits the minimal value that $z$ can take to be ${\mbox{O}}(a)$. 

\subsection{Renormalization and matching}
\label{sec:RENMAT}

In a renormalizable theory, like QCD, the scaling in the deep inelastic region is controlled by computable logarithmic corrections. Unfortunately, the local operators in eq.~(\ref{locopbil}) require a renormalization which is not simply multiplicative. In fact, the matrix elements~(\ref{locopbil}) also display UV power divergent mixing with lower dimensional (trace) operators that one needs to resolve to make both $A_n$ and $B_n$ form factors finite. In particular in order to be able to take the limit $P_z \to \infty$, necessary to eliminate the contamination from higher twists, one needs to make the $B_n$'s finite. The only renormalization considered in~\cite{Ji:2013dva,Lin:2017snn} was, however, the multiplicative ``matching condition'' which we now discuss.

The basic procedure, common to many of the approaches that have been following in a way or another the Ji paper~\cite{Ji:2013dva} is to start considering \begin{eqnarray}
&&{\tilde F} (x, P_z;\Lambda) = \frac {P_z}{2 \pi} \int_{- \infty}^{+\infty} dz \, e^{i x z P_z} \langle P |\phi(0) \phi(z)|P\rangle \Big{|}_\Lambda \, ,
\label{fund}
\end{eqnarray}
where $\Lambda$ is an UV cutoff. Renormalization is carried out by means of the so-called ``matching procedure'' which consists in writing
\begin{eqnarray}
 {\tilde F} (x, P_z; \Lambda) =  \int_x^{+\infty} \frac {d x'}{x'} Z (\frac {x}{x'}; \Lambda,\mu) F(x',P_z;\mu) \, ,
\end{eqnarray}
where $Z (\frac {x}{x'}; \Lambda,\mu )$ is a logarithmically divergent renormalization function (computed in perturbation theory - PT) which is needed to construct an UV finite $F(x,P_z; \mu)$.

We observe that the convolution property of the Mellin transform yields  
\begin{eqnarray}
\hspace{-.8cm}&&\int_{-\infty}^{+\infty} \!d x {\tilde F} (x, P_z; \Lambda) x^n\! =\! \int_{-\infty}^{+\infty} \!dx' {x'}^{\,n} Z(x'; \Lambda,\mu) \int_{-\infty}^{+\infty} \!dx \,x^n  F (x,P_z; \mu) \!\equiv \nonumber \\
\hspace{-.8cm}&&\equiv Z_n \left (\frac {\Lambda} {\mu} \right) \int_{-\infty}^{+\infty} \!dx\, x^n F (x,P_z; \mu)  \, , \label{mellconv}
\end{eqnarray}
implying that the moments of $\tilde F$ is are multiplicatively renormalized, independently from each other.

Eq.~(\ref{mellconv}) becomes a relation involving the moments of the physical PDF after sending $P_z\to \infty$. Taking this limit on the lattice is, however, not possible as we now show.

Eq.~(\ref{fund}) is a FT, the inverse of which reads 
\begin{eqnarray}
\langle P |\phi(0) \phi({z})|P\rangle \Big{|}_\Lambda = \int_{-\infty}^{+\infty} d x\, e^{- i x {z} P_z} {\tilde F}(x, P_z; \Lambda) \, . \label{genfun}
\end{eqnarray}
Taking the $n$-th derivative of~(\ref{genfun}) with respect to ${z}$ at ${z}=0$ gives 
\begin{eqnarray}
(-i)^n \int_{-\infty}^{+\infty} d x \, \, x^n {\tilde F}(x, P_z; \Lambda) = \frac {1}{(P_z)^n} \langle P |\phi(0) \frac {\partial^n \phi}{\partial z^n}(0)|P\rangle \Big{|}_\Lambda \, ,
\end{eqnarray}
which together with eq.~(\ref{mellconv}) implies
\begin{eqnarray}
&&\int_{-\infty}^{+\infty} dx \,x^n  F (x, P_z; \mu)  = \frac {(-i)^n} {Z_n (\Lambda/\mu)} \int_{-\infty}^{+\infty} d x \, \, x^n {\tilde F}(x, P_z; \Lambda) = \nonumber \\
&&= \frac {1}{(P_z)^n} \langle P | \frac {1} {Z_n (\Lambda/\mu)} \phi(0) \frac {\partial^n \phi}{\partial z^n}(0)|P\rangle \Big{|}_\Lambda \, . \label{finedellastoria}
\end{eqnarray}
The l.h.s.\ of eq.~(\ref{finedellastoria}) should yield the ``measurable, UV finite'' moments of the physical structure functions with $Z_n(\Lambda/\mu)$ the renormalization constants which would make the operators $\phi(0) \frac {\partial^n \phi}{\partial z^n}(0)$ finite. However, as already mentioned, these equal-point operators are not multiplicatively renormalizable due to the presence of power divergent divergent trace terms. They require subtractions and not simply a multiplicative renormalization.

\section{Reduced Ioffe-time distributions and perturbative subtraction}
\label{sec:RIT}

The difficulties outlined before, that prevent the direct calculation of the PDF on the lattice, affect also the strategy advocated in refs.~\cite{Orginos:2017kos,Zhang:2018ggy} where it is proposed to consider as a better UV behaved quantity the reduced Ioffe-time distribution~\cite{IOFFE} 
\beq
\mathfrak{M}(P_z\,z,z^2)=\frac{{\cal M}(-P_z\,z,-z^2)}{{\cal M}(0,-z^2)}
\label{ORGI}
\eeq
with ${\cal M}(-P_z\,z,-z^2)$ the bilocal of eq.~(\ref{lorbil}) evaluated at $\xi=(0,0,0,z)$. Since the ratio $\mathfrak{M}(P_z\,z,z^2)$ only differs from ${\cal M}(-P_z\,z,-z^2)$ by a ($z^2$-dependent) rescaling, the problem with power divergent moments is still present. Indeed, from the small $z$ OPE of the lattice regularized ratio~(\ref{ORGI}) in terms of ``continuum'' Wilson coefficients (say in the $\overline MS$ scheme) one can in principle extract the correct PDF moments. However, in order to directly construct the PDF from lattice data, 
one would need to take the FT of the quantity~(\ref{ORGI}). This FT will display power divergent moments irrespective of whether moments are defined as derivatives of the quasi-PDF with respect to $z$ (with fixed $P_z$) or of the pseudo-PDF with respect to $\nu=-P_z z$ (at fixed $z$). In the first case we are in the same situation as for the original Ji proposal (see our discussion in sect.~\ref{sec:RENMAT}). In the second, in order to take the derivatives with respect to $\nu=-P_z z$ at vanishing $z^2$ one must send $P_z\to \infty$, which is impossible in the lattice regularization. 

As a way to circumvent these problems the Authors of refs.~\cite{Orginos:2017kos} have proposed to subtract out the unwanted terms in PT. We illustrate their idea and the difficulties that go along with it with the help of the illuminating approach and notations of ref.~\cite{Radyushkin:2018nbf}.  

Using, say, dimensional regularization and the $\overline MS$ subtraction scheme, the regularized quasi-PDF $Q(y,P)$ can be related to the light-cone continuum PDF $f(y,\mu^2)$ by the perturbative formula~\cite{Zhang:2018ggy,Radyushkin:2018nbf}
\begin{eqnarray}
\hspace{-.6cm}&&Q(y,P)=f(y;\mu^2) -\frac{\alpha_s}{2\pi}C_F\int_0^1\frac{du}{u}f(\frac{y}{u}; \mu^2)\Big{[}B(u)\ln(\frac{\mu^2}{P^2})+C(u)\Big{]}+\nn\\
\hspace{-.6cm}&& + \frac{\alpha_s}{2\pi}C_F\int_{-1}^1 dx\, f(x;\mu^2) L(y,x) +{\mbox{O}}(P^{-2}) +{\mbox{O}}(\alpha_s^{2})
\, ,\label{RAD1}
\end{eqnarray}
where
\begin{eqnarray}
\hspace{-.6cm}&&L(y,x) =\frac{P}{2\pi}\int_0^1 du\, B(u) \int_{-\infty}^{+\infty} dz \,e^{-i(y-ux)zP} \ln (z^2P^2)\, .\label{RAD2}
\end{eqnarray}
The last term in eq.~(\ref{RAD1}) produces (unwanted) contributions in the $|y|>1$ region, responsible for UV power divergent moments. One can thus think of subtracting out by hand these terms writing
\begin{eqnarray}
\hspace{-.8cm}&&f(y;\mu^2)= \Big{[}{Q(y,P)}- \frac{\alpha_s}{2\pi}C_F\int_{-1}^1 dx\, f(x;\mu^2) L(y,x) \Big{]}+\nn\\
\hspace{-.8cm}&& +\frac{\alpha_s}{2\pi}C_F\int_0^1\frac{du}{u}f(\frac{y}{u};\mu^2)\Big{[}B(u)\ln(\frac{\mu^2}{P^2})+C(u)\Big{]}
+{\mbox{O}}(P^{-2}) +{\mbox{O}}(\alpha_s^{2})\, .\label{RAD3}
\end{eqnarray}
The difficulties posed by this procedure, which is widely used in actual simulations, are as follows. First of all we observe that the subtraction needs to be carried out before removing the cutoff. So all the above formulae should be looked at with this in mind. For instance, in lattice simulations eq.~(\ref{RAD1}) and the following should be rewritten by using the lattice regularization.

Secondly, although it is true that the term in square parenthesis has a smooth $P\to \infty$ limit, the ${\mbox{O}}(\alpha_s^{2})$ corrections don't and at small lattice spacings they will matter. Indeed, UV power divergences in moments are not eliminated but only pushed to higher orders in PT. 

Finally the very same PDF, $f(y;\mu^2)$, one is looking for appears in the r.h.s.\ of eq.~(\ref{RAD3}). In practice to leading order in $\alpha_s$ one replaces it with the lattice quantity $Q(y,P)$. The latter, however does not have the correct support properties. One thus needs to enforce them by hand. As a result non-localities are introduced. Then the question arises whether the moments of the PDF built in this way are the matrix elements of the renormalized local DIS operators~(\ref{locopbil}) one finds in the Bjorken limit. 

\subsection{Observation}
\label{sec:OBS}

We wish to end this section with an important observation. Many Authors (see among others~\cite{Ji:2017rah}) consider the argument about UV power divergent moments inconclusive on the basis of the observation that the lattice expression of the bilocal hadronic matrix element behaves as $\log |z|$ around $z=0$ and that (possible) power divergencies appear only if the moments of lattice PDF are computed~\footnote{In Appendix~B of ref.~\cite{Rossi:2017muf} an explicit example of a reasonably smooth function nevertheless displaying divergent moments is discussed.}, something that it is either simply dismissed as unnecessary or claimed to be allowed only after "matching". 

Both counter-arguments are, however, untenable. First of all, the calculation of the moments from the lattice PDF is crucial as one can claim to have computed the correct PDF only if the moments of the latter reproduce the matrix elements of the renormalized local DIS operators dominating in the light-cone limit that one measures in experiments. Secondly, although it is true that renormalization (matching) can transform the non-local quasi-distribution function into a well defined mathematical object (technically a ``distribution'', i.e.\ a singular function with integrable singularities), when naively differentiated, as one has to do for computing moments, one gets increasingly singular behaviours. As eq.~(\ref{mellconv}) shows, no linear relation of the matching type can improve the  situation. 

\section{PDF from current-current $T$-products?}
\label{sec:QIUMA}

As an alternative to the Ji strategy, the Authors of ref.~\cite{Ma:2017pxb} propose to compute directly the hadronic matrix element of the $T$-product of two currents on the lattice~\footnote{In a different context, a similar idea was put forward in ref.~\cite{Dawson:1997ic} to bypass the difficulties with the lower dimension operator mixing in the case of the construction of the lattice $d=6$ effective weak Hamiltonian.}
\beq
\sigma(\omega,\xi^2) = \langle P|T(J(0)J(\xi))|P \rangle \, ,
\label{TJJ}
\eeq
where in the notation of ref.~\cite{Ma:2017pxb} $\omega = P\cdot \xi$. The idea of ref.~\cite{Ma:2017pxb} is to use the OPE, valid for small $\xi^2$, and reexpress $\sigma$ in terms of the product of the physical PDF times a perturbatively computable kernel integrated over the Bjorken variable. More concretely in ref.~\cite{Ma:2017pxb} it is proposed to start with the expansion   
\beqn
&&\sigma(\omega,\xi^2) = \sum_{n}W_n(\xi^2;\mu^2)\, \xi^{\mu_1} \xi^{\mu_2} \ldots\xi^{\mu_n} \langle P|O_{\mu_1\mu_2 \ldots \mu_n}(0)|P \rangle \label{TJJF}
\eeqn
and, after using eq.~(\ref{locopbil}) with 
\beq
A_n(\mu^2)=\int \frac{dx}{x}  \, x^n f(x;\mu^2)\, ,
\label{AJ}
\eeq
to cast eq.~(\ref{TJJF}) in the form
\beq
\sigma(\omega,\xi^2) =\int \frac{dx}{x} f(x;\mu^2) K(x\omega,\xi^2,x^2;\mu^2)+{\mbox{O}}(\xi^2\Lambda_{QCD}^2)\, ,
\label{FINEQ}
\eeq
where (eq.~(14) of ref.~\cite{Ma:2017pxb})
\beqn
\hspace{-1.2cm}&&K(x\omega,\xi^2,x^2;\mu^2)=\nn\\
\hspace{-1.2cm}&&\quad=\sum_n x^n W_n(\xi^2;\mu^2) \, \xi^{\mu_1} \xi^{\mu_2} \ldots\xi^{\mu_n} (P_{\mu_1} P_{\mu_2}\ldots P_{\mu_n} + traces)\, .
\label{KW}
\eeqn
The Authors conclude that, to the extent that $K$ is known in PT~\footnote{For instance, they find $K^q_q(x\omega,\xi^2,0;\mu)=2x\omega \exp{(ix\omega)}$.}, $f(x;\mu^2)$ can be obtained as the one-dimensional FT  (eq.~(24) of ref.~\cite{Ma:2017pxb})
\beqn
\hspace{-1.2cm}&&  \frac{1}{4\pi}
\int \frac{d\omega}{\omega} \, {\mbox e}^{-ix\omega}\sigma(\omega,\xi^2)=f(x;\mu^2) + {\mbox{O}}(\xi^2\Lambda_{QCD}^2)\, ,
\label{FX}
\eeqn
if lattice data are inserted for $\sigma$. The trouble with this equation is that it is sensitive to contributions from higher twists (${\mbox{O}}(\xi^2\Lambda_{QCD}^2)$ terms). To give higher twists a vanishing weight one should take, besides $\xi^0=0$, also the limit $\xi^3=z \to 0$ in order to maintain the Euclidean constraint $\xi^2\to 0$. If one does so, however, to keep the integration variable $\omega$ fixed, one needs to send $P_z\to \infty$ as $z\to 0$. But this is impossible as the accessible values of $P_z$ are limited by the lattice UV cutoff. In this respect we have here a situation similar to the one we encountered at the end of sect.~\ref{sec:EM}.

Hence even the approach developed in ref.~\cite{Ma:2017pxb} is unable to circumvent the criticism raised in~\cite{Rossi:2017muf} concerning the possibility of directly computing the PDF's in lattice simulations. 

Euclidean lattice data can instead give access to PDF moments. Moments can be extracted by numerically fitting the singular $\xi$ dependence of the current-current $T$-product~\cite{Ma:2017pxb} or of the bilocal~\cite{Karpie:2018zaz}, similarly to what was proposed to do in ref.~\cite{Dawson:1997ic} to get around the renormalization problems associated with the construction of the lattice effective weak Hamiltonian. 

\section{Conclusions}
\label{sec:CONCL}

In this note we have rediscussed the feasibility of the proposal of directly extracting PDF's from lattice simulations. Unfortunately there is still a missing ingredient in this program, related to the problem of subtracting power divergent trace terms. Although individual PDF moments can be extracted from lattice data, at this time neither the initial Ji idea of employing the bilocal operator~\cite{Ji:2013dva}, nor the reduced Ioffe-time distributions~\cite{Orginos:2017kos,Zhang:2018ggy}, or the direct use of the current-current $T$-product~\cite{Ma:2017pxb} allow to access the full PDF in lattice simulations. 

\vspace{.4cm}
{\bf Acknowledgments} - We wish to thank G.\ Martinelli and S.\ Zafeiropoulos for useful discussions.


\thebibliography{99}

%\cite{Ji:2013dva}
\bibitem{Ji:2013dva}
  X.~Ji,
  %``Parton Physics on a Euclidean Lattice,''
  Phys.\ Rev.\ Lett.\  {\bf 110} (2013) 262002.
%  [arXiv:1305.1539 [hep-ph]].
  %%CITATION = ARXIV:1305.1539;%%
  %50 citations counted in INSPIRE as of 13 Jul 2015

%\cite{Lin:2017snn}
\bibitem{Lin:2017snn}
  H.~W.~Lin {\it et al.},
  %``Parton distributions and lattice QCD calculations: a community white paper,''
  arXiv:1711.07916 [hep-ph].
  %%CITATION = ARXIV:1711.07916;%%
  %5 citations counted in INSPIRE as of 23 Jan 2018

%\cite{Brandt:1972nw}
\bibitem{Brandt:1972nw}
  R.~A.~Brandt and G.~Preparata,
  %``The light cone and photon-hadron interactions,''
  Fortsch.\ Phys.\  {\bf 20} (1972) 571.
  %%CITATION = FPYKA,20,571;%%
  %4 citations counted in INSPIRE as of 20 aožt 2015

%\cite{Rossi:2017muf}
\bibitem{Rossi:2017muf}
  G.~C.~Rossi and M.~Testa,
  %``Note on lattice regularization and equal-time correlators for parton distribution functions,''
  Phys.\ Rev.\ D {\bf 96} (2017) no.1,  014507.
 % doi:10.1103/PhysRevD.96.014507
%  [arXiv:1706.04428 [hep-lat]].
  %%CITATION = doi:10.1103/PhysRevD.96.014507;%%
  %13 citations counted in INSPIRE as of 23 Jan 2018

%\cite{Orginos:2017kos}
\bibitem{Orginos:2017kos}
  K.~Orginos, A.~Radyushkin, J.~Karpie and S.~Zafeiropoulos,
  %``Lattice QCD exploration of parton pseudo-distribution functions,''
  Phys.\ Rev.\ D {\bf 96} (2017) no.9,  094503.
  %doi:10.1103/PhysRevD.96.094503  [arXiv:1706.05373 [hep-ph]].
  %%CITATION = doi:10.1103/PhysRevD.96.094503;%%
  %38 citations counted in INSPIRE as of 27 Apr 2018

  %\cite{Zhang:2018ggy}
\bibitem{Zhang:2018ggy}
  J.~H.~Zhang, J.~W.~Chen and C.~Monahan,
  %``Parton distribution functions from reduced Ioffe-time distributions,''
  Phys.\ Rev.\ D {\bf 97} (2018) no.7,  074508.
%  doi:10.1103/PhysRevD.97.074508 [arXiv:1801.03023 [hep-ph]].
  %%CITATION = doi:10.1103/PhysRevD.97.074508;%%
  %7 citations counted in INSPIRE as of 15 May 2018

  %\cite{Ma:2017pxb}
\bibitem{Ma:2017pxb}
  Y.~Q.~Ma and J.~W.~Qiu,
  %``Exploring Partonic Structure of Hadrons Using ab initio Lattice QCD Calculations,''
  Phys.\ Rev.\ Lett.\  {\bf 120} (2018) no.2,  022003.
 % doi:10.1103/PhysRevLett.120.022003
%  [arXiv:1709.03018 [hep-ph]].
  %%CITATION = doi:10.1103/PhysRevLett.120.022003;%%
  %10 citations counted in INSPIRE as of 23 Jan 2018

\bibitem{IOFFE}
B.~L.~Ioffe, 
%Space-time picture of photon and neutrino scattering and electroproduction cross-section asymptotics, 
Phys.\ Lett.\ {\bf 30B} (1969) 123.

%\cite{Radyushkin:2017cyf}
\bibitem{Radyushkin:2017cyf}
  A.~V.~Radyushkin,
  %``Quasi-parton distribution functions, momentum distributions, and pseudo-parton distribution functions,''
  Phys.\ Rev.\ D {\bf 96} (2017) no.3,  034025.
 % doi:10.1103/PhysRevD.96.034025 [arXiv:1705.01488 [hep-ph]].
  %%CITATION = doi:10.1103/PhysRevD.96.034025;%%
  %46 citations counted in INSPIRE as of 17 Aug 2018

%\cite{Radyushkin:2018nbf}
\bibitem{Radyushkin:2018nbf}
  A.~V.~Radyushkin,
  %``Structure of parton quasi-distributions and their moments,''
  arXiv:1807.07509 [hep-ph].
  %%CITATION = ARXIV:1807.07509;%%
  %2 citations counted in INSPIRE as of 13 Aug 2018

%\cite{Ji:2017rah}
\bibitem{Ji:2017rah}
  X.~Ji, J.~H.~Zhang and Y.~Zhao,
  %``More On Large-Momentum Effective Theory Approach to Parton Physics,''
  Nucl.\ Phys.\ B {\bf 924} (2017) 366.
%  doi:10.1016/j.nuclphysb.2017.09.001[arXiv:1706.07416 [hep-ph]].
  %%CITATION = doi:10.1016/j.nuclphysb.2017.09.001;%%
  %26 citations counted in INSPIRE as of 01 Jun 2018

%\cite{Dawson:1997ic}
\bibitem{Dawson:1997ic}
 C.~Dawson, G.~Martinelli, G.~C.~Rossi, C.~T.~Sachrajda, S.~R.~Sharpe, M.~Talevi and M.~Testa,
  %``New lattice approaches to the delta I = 1/2 rule,''
  Nucl.\ Phys.\ B {\bf 514} (1998) 313.
  %doi:10.1016/S0550-3213(97)00756-6
%  [hep-lat/9707009].
  %%CITATION = doi:10.1016/S0550-3213(97)00756-6;%%
  
  %\cite{Karpie:2018zaz}
\bibitem{Karpie:2018zaz}
  J.~Karpie, K.~Orginos and S.~Zafeiropoulos,
  %``Moments of Ioffe time parton distribution functions from non-local matrix elements,''
  arXiv:1807.10933 [hep-lat].
\end{document}